\documentclass[reprint,aip,jcp,amsmath,amssymb]{revtex4-1}
\usepackage[utf8]{inputenc}
\usepackage{amsmath}
\usepackage{graphicx}   
\usepackage{dcolumn}    
\usepackage{bm}         
\usepackage{hyperref}
\usepackage{textcomp}
\usepackage{amsmath}
\usepackage{array,multirow}

\newcommand{\etf}{e^{-i \hat H t/\hbar}}
\newcommand{\etb}{e^{i \hat H t/\hbar}}

\newcommand{\dfq}{\delta[f(\bq)]}

\newcommand{\dd}[2]{\frac{d #1}{d #2}}

\newcommand{\ddp}[2]{\frac{\partial #1}{\partial #2}}

\newcommand{\smiNz}{\sum_{i=0}^{N-1}}

\newcommand{\smjNz}{\sum_{j=0}^{N-1}}

\newcommand{\ltti}{\lim_{t\to \infty}}

\newcommand{\shortt}{$t\to 0_+$}
\newcommand{\longt}{$t\to \infty$}

\newcommand{\tr}{ {\rm Tr} }
\newcommand{\qdd}{q^\ddag}

\newcommand{\Qrb}{Q_{\rm r}(\beta)}

\newcommand{\no}{\nonumber}

\newcommand{\betaN}{\beta_N}
\newcommand{\ola}{\overleftarrow}
\newcommand{\ora}{\overrightarrow}

\newcommand{\bp}{ {\bf p} }

\newcommand{\bq}{ {\bf q} }

\newcommand{\fq}{ f({\bf q}) }

\newcommand{\tphN}{\frac{1}{(2\pi\hbar)^N}}

\newcommand{\inti}{\int_{-\infty}^{\infty}}
\newcommand{\eqr}[1]{Eq.~\eqref{eq:#1}}
\newcommand{\eqsr}[2]{Eqs.~\eqref{eq:#1} and \eqref{eq:#2}}
\newcommand{\eql}[1]{\label{eq:#1}}
\newcommand{\figr}[1]{Fig.~\ref{fig:#1}}
\newcommand{\figl}[1]{\label{fig:#1}}

\newcommand{\tar}{\tilde \alpha_{\rm RP}}
\begin{document}

\title{Should Thermostatted Ring Polymer Molecular Dynamics be used to calculate thermal reaction rates?} 
\author{Timothy J.~H.~Hele\footnote{Corresponding author: tjhh2@cam.ac.uk}}
\affiliation{Department of Chemistry, University of Cambridge, Lensfield Road, Cambridge, CB2 1EW, UK.}
\author{Yury V.~Suleimanov}
\affiliation{Computation-based Science and Technology Research Center, Cyprus Institute, 20 Kavafi Str., Nicosia 2121, Cyprus}
\affiliation{Department of Chemical Engineering, Massachusetts Institute of Technology, 77 Massachusetts Ave., Cambridge, Massachusetts 02139, United States}
\date{\today}

\begin{abstract}
We apply Thermostatted Ring Polymer Molecular Dynamics (TRPMD), a recently-proposed approximate quantum dynamics method, to the computation of thermal reaction rates. Its short-time Transition-State Theory (TST) limit is identical to rigorous Quantum Transition-State Theory, and we find that its long-time limit is independent of the location of the dividing surface. TRPMD rate theory is then applied to one-dimensional model systems, the atom-diatom bimolecular reactions H+H$_2$, D+MuH and F+H$_2$, and the prototypical polyatomic reaction H+CH$_4$. Above the crossover temperature, the TRPMD rate is virtually invariant to the strength of the friction applied to the internal ring-polymer normal modes, and beneath the crossover temperature the TRPMD rate generally decreases with increasing friction, in agreement with the predictions of Kramers theory. We therefore find that TRPMD is approximately equal to, or less accurate than, Ring Polymer Molecular Dynamics (RPMD) for symmetric reactions, and for certain asymmetric systems and friction parameters closer to the quantum result, providing a basis for further assessment of the accuracy of this method. 
\emph{Copyright (2015) American Institute of Physics. This article may be downloaded for personal use only. Any other use requires prior permission of the author and the American Institute of Physics. The following article appeared in J.~Chem.~Phys., \textbf{143} (2015), 074107, and may be found at  http://dx.doi.org/10.1063/1.4928599.}
\end{abstract}

\maketitle 

\section{Introduction}
The accurate computation of thermal quantum rates is a major challenge in theoretical chemistry, as a purely classical description of the kinetics fails to capture zero-point energy, tunnelling, and phase effects\cite{pol05,han90}. Exact solutions using correlation functions, developed by Yamamoto, Miller, and others\cite{yam59, mil74, mil83, mil93} are only tractable for small or model systems, as the difficulty of computation scales exponentially with the size of the system.

Consequently, numerous approximate treatments have been developed, which can be broadly classed as those seeking an accurate description of the quantum statistics without direct calculation of the dynamics, and those which also seek to use an approximate quantum dynamics. Methods in the first category include instanton theory\cite{mil75,cha75,aff81,mil97,mil08,and09,ric09, alt11,kaw14,rom11,zha14}, ``quantum instanton''\cite{van05,mil03}, and various transition-state theory (TST) approaches\cite{eyr35,eyr35rev,wig38,tru96,tru80,vot89rig}. Of many approximate quantum dynamics methods, particularly successful ones include the linearized semiclassical initial-value representation (LSC-IVR) \cite{liu09,liu15,shi03}, centroid molecular dynamics (CMD)\cite{cao93,*cao94_1,*cao94_2,*cao94_3,*cao94,*cao94_5, vot93,vot96,shi02,hon06,jan99cmd,jan99pi}, and ring polymer molecular dynamics (RPMD)\cite{man04,man05che,man05ref,hab13}.

RPMD has been very successful for the computation of thermal quantum rates in condensed-phase processes, due to the possibility of implementation in complex systems such as (proton-coupled) electron transfer reaction dynamics or enzyme catalysis,\cite{boe11,kre13,men11,men14} and especially in small gas-phase systems\cite{hab13,col09,col10,sul11,sul12,sul14,per14stress,lix13,lix13kie,sul13hh2,all13,per12,per14,lix14kie,esp14,gon14,lix14,sul15,hic15} where comparison with exact quantum rates and experimental data has demonstrated that RPMD rate theory is a consistent and reliable approach with a high level of accuracy. These numerical results have shown that RPMD rate theory is exact in the high-temperature limit (which can also be shown algebraically\cite{hab13}), reliable at intermediate temperatures, and more accurate than other approximate methods in the deep tunnelling regime (see \eqr{xover} below), where it is within a factor of 2--3 of the exact quantum result. RPMD also captures zero-point energy effects,\cite{per12} and provides very accurate estimates for barrierless reactions\cite{lix14,sul14}. It has been found to systematically overestimate thermal rates for asymmetric reactions and underestimate them for symmetric (and quasisymmetric) reactions in the deep tunnelling regime (Note that zero-point energy effects along the reaction coordinate must be taken into account when assigning the reaction symmetry.)\cite{ric09,sul13hh2} Recently a general code for RPMD calculations (RPMDrate) has been developed.\cite{sul13}

Another appealing feature of RPMD rate theory is its rigorous independence to the location of the dividing surface between products and reactants\cite{man05ref}, a property shared by classical rate theory and the exact quantum rate\cite{man05ref}, but not by many transition-state theory approaches. The \shortt, TST limit of RPMD (RPMD-TST) is identical to true QTST: the instantaneous thermal quantum flux through a position-space dividing surface which is equal to the exact quantum rate in the absence of recrossing\cite{hel13,alt13,hel13unique}. A corollary of this is that RPMD will be exact for a parabolic barrier (where there is no recrossing of the optimal dividing surface by RPMD dynamics or quantum dynamics, and QTST is therefore also exact)\cite{alt13}. 

When the centroid is used as the dividing surface (see \eqr{cen} below), RPMD-TST reduces to the earlier theory of centroid-TST\cite{gil87,gil87hyd,vot89rig,vot93}, which is a good approximation for symmetric barriers but significantly overestimates the rate for asymmetric barriers at low temperatures\cite{man05ref,gev01,col09}. This effect is attributable to the centroid being a poor dividing surface beneath the `crossover' temperature into deep tunnelling\cite{ric09}. In this `deep tunnelling' regime, RPMD-TST has a close relationship to semiclassical ``Im F'' instanton theory\cite{ric09,ric12}, which has been very successful for calculating rates beneath the crossover temperature, though has no first-principles derivation\cite{alt11} and was recently shown to be less accurate than QTST when applied to realistic multidimensional reactions\cite{zha14com}. 

Very recently, both CMD and RPMD have been obtained from the exact quantum Kubo-transformed\cite{kub57} time-correlation function (with explicit error terms) via a Boltzmann-conserving ``Matsubara dynamics''\cite{hel15,hel15rel} which considers evolution of the low-frequency, smooth ``Matsubara'' modes of the path integral\cite{mat55}. Matsubara dynamics suffers from the sign problem and is not presently amenable to computation on large systems. However, by taking a mean-field approximation to the centroid dynamics, such that fluctuations around the centroid are discarded, one obtains CMD.\cite{hel15rel} Alternatively, if the momentum contour is moved into the complex plane in order to make the quantum Boltzmann distribution real, a complex Liouvillian arises, the imaginary part of which only affects the higher, non-centroid, normal modes. Discarding the imaginary Liouvillian leads to spurious springs in the dynamics and gives RPMD.\cite{hel15rel} Consequently, RPMD will be a reasonable approximation to Matsubara dynamics, provided that the timescale over which the resultant dynamics is required (the timescale of `falling off' the barrier in rate theory) is shorter than the timescale over which the springs `contaminate' the dynamics of interest (in rate theory, this is usually coupling of the springs in the higher normal modes to the motion of the centroid dividing surface via anharmonicity in the potential).

Both RPMD and CMD are inaccurate for the computation of multidimensional spectra: the neglect of fluctuations in CMD leads to the ``curvature problem'' where the spectrum is red-shifted and broadened, whereas in RPMD the springs couple to the external potential leading to ``spurious resonances''\cite{wit09,iva10}. Recently, this problem has been solved by attaching a Langevin thermostat\cite{bus07} to the internal modes of the ring polymer\cite{ros14} (which had previously been used for the computation of statistical properties\cite{cer10}), and the resulting Thermostatted RPMD (TRPMD) had neither the curvature nor resonance problem. 

The success of RPMD for rate calculation, and the attachment of a thermostat for improving its computation of spectra, naturally motivates studying whether TRPMD will be superior for the computation of thermal quantum rates to RPMD (and other approximate theories)\cite{per14stress,ros14}, which this article investigates. Given that RPMD is one of the most accurate approximate methods for systems where the quantum rates are available for comparison, further improvements would be of considerable benefit to the field. 

We firstly review TRPMD dynamics in section \ref{ssec:trpmd}, followed by developing TRPMD rate theory in section~\ref{ssec:rate}. To predict the behaviour of the RPMD rate compared to the TRPMD rate, we apply one-dimensional Kramers theory\cite{kra40} to the ring-polymer potential energy surface in section~\ref{ssec:kram}. Numerical results in section~\ref{sec:num} apply TRPMD to the symmetric and asymmetric Eckart barriers followed by representative bimolecular reactions: H+H$_2$ (symmetric), D+MuH (quasisymmetrical), H+CH$_4$ (prototypical polyatomic reaction) and F+H$_2$ (asymmetric and highly anharmonic). Conclusions and avenues for further research are presented in section~\ref{sec:con}.

%

\section{Theory}
\label{sec:theory}
\subsection{Thermostatted Ring Polymer Molecular Dynamics}
\label{ssec:trpmd}
For simplicity we consider a one-dimensional system $(F=1)$ with position $q$ and associated momentum $p$ at inverse temperature $\beta = 1/k_{\rm B}T$, where the $N$-bead ring-polymer Hamiltonian is\cite{fey65,man04} 
\begin{align}
 H_{N}(\bp,\bq) = \smiNz \frac{p_i^2}{2m} + U_N(\bq) \eql{rpham}
\end{align}
with the ring-polymer potential
\begin{align}
 U_N(\bq) =  \smiNz \tfrac{1}{2} m \omega_N^2 (q_i-q_{i-1})^2 + V(q_i) \eql{rppot}
\end{align}
and the frequency of the ring-polymer springs $\omega_N = 1/\betaN\hbar$, where $\betaN \equiv \beta/N$. Generalization to further dimensions follows immediately, and merely requires more indices.\cite{ros14}

The ring polymer is time-evolved by propagating stochastic trajectories using TRPMD dynamics\cite{cer10,ros14},
\begin{align}
 \dot \bp = & -\nabla_{\bq} U_N(\bq) - \bm{\Gamma} \bp + \sqrt{\frac{2m\bm{\Gamma}}{\betaN}} \bm{\xi}(t) \eql{dynp} \\
 \dot \bq = & \frac{1}{m} \bp \eql{dynq}
\end{align}
where $\bq\equiv (q_0,\ldots,q_{N-1})$ is the vector of bead positions and $\bp$ the vector of bead momenta, with $\nabla_{\bq}$ the grad operator in position-space, $\bm{\xi}(t)$ a vector of $N$ uniform Gaussian deviates with zero mean and unit variance, and $\bm{\Gamma}$ the $N\times N$ positive semi-definite friction matrix\cite{ros14}. 

The Fokker-Planck operator corresponding to the TRPMD dynamics in \eqsr{dynp}{dynq} is\cite{zwa01} 
\begin{align}
 \mathcal{A}_N = & - \frac{\bp}{m} \cdot \nabla_\bq + U_N(\bq) \ola \nabla_\bq \cdot \ora \nabla_\bp \no \\
 & + \nabla_{\bp}\cdot \bm{\Gamma}\cdot \bp + \frac{m}{\betaN} \nabla_\bp \cdot \bm{\Gamma} \cdot \nabla_\bp \eql{fpo}
\end{align}
(where the arrows correspond to the direction in which the derivative acts\cite{hel15}) and 
for any $\bm{\Gamma}$, TRPMD dynamics will conserve the quantum Boltzmann distribution ($\mathcal{A}_N e^{-\betaN H_N(\bp,\bq)} = 0$), a feature shared by RPMD and CMD but not some other approximate methods such as LSC-IVR\cite{hel15,hel15rel,liu15,liu09}. We then show in appendix~\ref{ap:db} that TRPMD obeys detailed balance, such that the TRPMD correlation function is invariant to swapping the operators at zero time and finite time, and changing the sign of the momenta.

The time-evolution of an observable is given by the adjoint of \eqr{fpo},\cite{ros14,zwa01}
\begin{align}
 \mathcal{A}_N^\dag = & \frac{\bp}{m} \cdot \nabla_\bq - U_N(\bq) \ola \nabla_\bq \cdot \ora \nabla_\bp \no\\
 & - \bp \cdot\bm{\Gamma}\cdot \nabla_{\bp} + \frac{m}{\betaN} \nabla_\bp \cdot \bm{\Gamma} \cdot \nabla_\bp \eql{fpa}.
\end{align}
In the zero-friction limit, $\bm{\Gamma} = \bm{0}$ and $\mathcal{A}_N^\dag = \mathcal{L}_N^{\dag}$, where $\mathcal{L_N}^{\dag}$ is the adjoint of the Liouvillian corresponding to deterministic ring-polymer trajectories\cite{hel15rel}. 

\subsection{TRPMD rate theory}
\label{ssec:rate}
We assume the standard depiction of rate dynamics, with a thermal distribution of reactants and a dividing surface in position space. In what follows we assume scattering dynamics, with the potential tending to a constant value at large separation of products and reactants. The methodology is then immediately applicable to condensed phase systems subject to the usual caveat that there is sufficient separation of timescales between reaction and equilibration.\cite{hel13,cha78} 

The exact quantum rate can be formally given as the long-time limit of the flux-side time-correlation function\cite{yam59,mil74,mil83}
\begin{align}
 k_{\rm QM}(\beta) = \lim_{t\to \infty} \frac{c_{\rm fs}^{\rm QM}(t)}{\Qrb}
\end{align}
where $\Qrb$ is the partition function in the reactant region and\footnote{We note that the flux-side function in the gas phase was originally derived as an asymmetric-split trace\cite{mil74} and later as a symmetric-split trace\cite{mil83}, both of which are equivalent to the Kubo transformed trace given here in the \longt\ limit.} 
\begin{align}
 c_{\rm fs}^{\rm QM}(t) = \frac{1}{\beta} \int_{0}^{\beta} d\sigma\ \tr \left[ e^{-(\beta - \sigma)\hat H} \hat F e^{-\sigma \hat H} \etb \hat h \etf \right] \eql{cfsqm}
\end{align}
with $\hat F$ and $\hat h$ the quantum flux and side operators respectively, and $\hat H$ the Hamiltonian for the system. The quantum rate can equivalently be given as minus the long-time limit of the time-derivative of the side-side correlation function, or the integral over the flux-flux correlation function\cite{mil83}.

The TRPMD side-side correlation function is 
\begin{align}
 C_{\rm ss}^{\rm TRPMD}(t) = & \tphN \int d\bp \int d\bq\ \no\\
 & \times e^{-\betaN H_N(\bp,\bq)} h[f(\bq)] h[f(\bq_t)] \eql{css}
\end{align}
where $\int d\bq \equiv \inti dq_0 \inti dq_1 \ldots \inti dq_{N-1}$ and likewise for $\int d\bp$, and $\bq_t \equiv \bq_t(\bp,\bq,t)$ is obtained by evolution of $(\bp,\bq)$ for time $t$ with TRPMD dynamics. The ring polymer reaction co-ordinate $\fq$ is defined such that the dividing surface is at $\fq = 0$, and that $\fq > 0$ corresponds to products and $\fq < 0$ to reactants. 

Direct differentiation of the side-side correlation function using the Fokker--Planck operator in \eqr{fpo} yields the TRPMD flux-side time-correlation function
\begin{align}
 C_{\rm fs}^{\rm TRPMD}(t) =&  - \dd{}{t}C_{\rm ss}^{\rm TRPMD}(t) \eql{cssfs} \\
 = \tphN & \int d\bp \int d\bq\ e^{-\betaN H_N(\bp,\bq)}\no\\
  & \times \dfq S_N(\bp,\bq) h[f(\bq_t)] \eql{cfs}
\end{align}
where $S_N(\bp,\bq)$ is the flux perpendicular to $\fq$ at time $t=0$,
\begin{align}
 S_N(\bp,\bq) = \smiNz \ddp{\fq}{q_i} \frac{p_i}{m}.
\end{align}
We approximate the long-time limit of the quantum flux-side time-correlation function in \eqr{cfsqm} as the long-time limit of the TRPMD flux-side time-correlation function in \eqr{cfs}, leading to the TRPMD approximation to the quantum rate as
\begin{align}
 k_{\rm TRPMD}(\beta) = \lim_{t\to\infty}\frac{C_{\rm fs}^{\rm TRPMD}(t)}{\Qrb}. \eql{trpmdrate}
\end{align}

The flux-side time-correlation function \eqr{cfs} will decay from an initial TST (\shortt) value to a plateau, which (for a gas-phase scattering reaction with no friction on motion out of the reactant or product channel) will extend to infinity. For condensed-phase reactions (and gas-phase reactions with friction in exit channels) a rate is defined provided that there is sufficient separation of timescales between reaction and equilibration to define a plateau in $C_{\rm fs}^{\rm TRPMD}(t)$,\cite{cha78} which at very long times (of the order $k_{\rm TRPMD}(\beta)^{-1}$ for a unimolecular reaction) tends to zero\footnote{See e.g.\ Fig.~5 of Ref.~{\protect \onlinecite{sul12}}.}. 

Further differentiation of the flux-side time-correlation function (with the adjoint of the Fokker-Planck operator in \eqr{fpa}) yields the TRPMD flux-flux correlation function
\begin{align}
  C_{\rm ff}^{\rm TRPMD}(t) = & \tphN \int d\bp \int d\bq \ e^{-\betaN H_N(\bp,\bq)} \no\\
  & \times \dfq S_N(\bp,\bq) \delta[f(\bq_t)] S_N(\bp_t,\bq_t) \eql{cff}
\end{align}
which, by construction, must be zero in the plateau region, during which no trajectories recross the dividing surface. 

Like RPMD rate theory, TRPMD has the appealing feature that its short-time (TST) limit is identical to true Quantum Transition-State Theory (QTST), as can be observed by applying the short-time limit of the Fokker-Planck propagator $e^{\mathcal{A}_N^\dag t}$ to $\fq$, yielding\cite{ros14} 
\begin{align}
 \lim_{t\to 0_+} \frac{C_{\rm fs}^{\rm TRPMD}(t)}{\Qrb} = k^\ddag_{\rm QM} (\beta) \eql{qtst}
\end{align}
where $k^\ddag_{\rm QM}(\beta)$ is the QTST rate\cite{hel13,alt13,hel13unique,hel14PhD}. In Appendix~\ref{ap:ind} we then show that the TRPMD rate in \eqr{trpmdrate} is rigorously independent of the location of the dividing surface.  Consequently, the TRPMD rate will equal the exact quantum rate in the absence of recrossing of the optimal dividing surface (and those orthogonal to it in path-integral space) by either the exact quantum or TRPMD dynamics.\cite{alt13} We also note that \eqr{qtst} holds regardless of the value of the friction matrix $\bm{\Gamma}$ and that recrossing of individual (stochastic) trajectories can only reduce the TRPMD rate from the QTST value, and hence QTST is an upper bound to the long-time TRPMD rate.

In the following calculations we use a friction matrix which corresponds to damping of the free ring polymer vibrational frequencies, and which has been used in previous studies of TRPMD for spectra.\cite{ros14,ros14com} For an orthogonal transformation matrix $\mathbf{T}$ such that
\begin{align}
 \mathbf{T}^T \mathbf{K} \mathbf{T} = m \bm{\Omega}^2
\end{align}
where $\mathbf{K}$ is the spring matrix in \eqr{rppot} and $\bm{\Omega}_{ij} = 2\delta_{ij} \sin(j\pi/N)/\betaN\hbar$, the friction matrix is given by
\begin{align}
 \mathbf{\Gamma} = 2\lambda \mathbf{T} \bm{\Omega} \mathbf{T}^T.
\end{align}
Here $\lambda$ is an adjustable parameter, with $\lambda = 1$ giving critical damping of the free ring polymer vibrations, $\lambda = 0.5$ corresponding to optimal sampling of the free ring polymer potential energy, and $\lambda = 0$ corresponding to zero friction (i.e.\ RPMD).\cite{ros14,cer10} A crucial consequence of this choice of friction matrix is that the centroid of the ring polymer is unthermostatted, and the short-time error of TRPMD from exact quantum dynamics is therefore $\mathcal{O}(t^7)$, the same as RPMD.\cite{ros14,bra06}

\subsection{Relation to Kramers Theory}
\label{ssec:kram}

To provide a qualitative description of the effect of friction on the TRPMD transmission coefficient, we apply classical Kramers theory\cite{kra40} in the extended $NF$-dimensional ring polymer space, governed by dynamics on the (temperature-dependent) ring-polymer potential energy surface in \eqr{rppot}. 
Since the short-time limit of TRPMD rate theory is equal to QTST, and its long-time limit invariant to the location of the dividing surface, TRPMD will give the QTST rate through the optimal dividing surface (defined as the surface which minimises $k^{\ddag}_{\rm QM}(\beta)$)\cite{ric09}, weighted by any recrossings of that dividing surface by the respective dynamics. 
We express this using the Bennett-Chandler factorization\cite{fre02},
\begin{align}
 k_{\rm TRPMD}(\beta) = & k^{\ddag*}_{\rm QM}(\beta) \lim_{t\to\infty} \kappa^*_{\rm TRPMD}(t) \eql{bc}
\end{align}
where $k^{\ddag *}_{\rm QM}(\beta)$ is the QTST rate, the asterisk denotes that the optimal dividing surface $f^*(\bq)$ is used and the TRPMD transmission coefficient is
\begin{widetext}
\begin{align}
 \kappa^*_{\rm TRPMD}(t) = \frac{\int d\bp \int d\bq\ e^{-\betaN H_N(\bp,\bq)} \delta[f^*(\bq)] S_N^*(\bp,\bq) h[f^*(\bq_t)] }{\int d\bp \int d\bq\ e^{-\betaN H_N(\bp,\bq)} \delta[f^*(\bq)] S_N^*(\bp,\bq) h[S_N^*(\bp,\bq)]} \eql{transcoef}
\end{align}
\end{widetext}
with analogous expressions to \eqsr{bc}{transcoef} for RPMD. To examine the explicit effect of friction on the TRPMD rate we define the ratio
\begin{align}
 \chi_{\lambda}(\beta) = \frac{k_{\rm TRPMD}(\beta)}{k_{\rm RPMD}(\beta)} \eql{chi}
\end{align}
and from \eqr{bc}
\begin{align}
 \chi_{\lambda}(\beta) = \lim_{t\to\infty} \frac{\kappa^*_{\rm TRPMD}(t)}{\kappa^*_{\rm RPMD}(t)}.
\end{align}
We then assume that the recrossing dynamics is dominated by one-dimensional motion through a parabolic saddle point on the ring-polymer potential energy surface, in which case the TRPMD transmission coefficient can be approximated by the Kramers expression\cite{kra40,nit06,fre02,pol86}
\begin{align}
\ltti \kappa^*_{\rm TRPMD}(t) \simeq \sqrt{1+\alpha_{\rm RP}^2} - \alpha_{\rm RP} \eql{klam}
\end{align}
where formally $\alpha_{\rm RP} = \gamma_{\rm RP}/2\omega_{\rm RP}$, with $\gamma_{\rm RP}$ the friction along the reaction co-ordinate and $\omega_{\rm RP}$ the barrier frequency in ring-polymer space. For a general $F$-dimensional system finding $f^*(\bq)$ and thereby computing $\gamma_{\rm RP}$ and $\omega_{\rm RP}$ is largely intractable. However, we expect $\gamma_{\rm RP} \propto \lambda$, and therefore define $\tilde \alpha_{\rm RP} = \alpha_{\rm RP}/\lambda$ where the dimensionless parameter $\tilde \alpha_{\rm RP}$ is expected to be independent of $\lambda$ for a given system and temperature, and represents the sensitivity of the TRPMD rate to friction. We further approximate that there is minimal recrossing of the optimal dividing surface by the (unthermostatted) ring polymer trajectories such that $\ltti \kappa^*_{\rm RPMD}(t) \simeq 1$,\footnote{In practice, both the location of the dividing surface and the degree of recrossing through it are very difficult to determine numerically, though Richardson and Althorpe{\protect \cite{ric09}} found that for the symmetric and asymmetric Eckart barrier at almost twice the inverse crossover temperature ($\beta\hbar\omega_b = 12$) fewer than 20\% of trajectories recrossed.} leading to
\begin{align}
 \chi_{\lambda}(\beta) \simeq \sqrt{1+\lambda^2 \tilde \alpha_{\rm RP}^2} - \lambda \tilde \alpha_{\rm RP}.\eql{alpharel}
\end{align}

Equation~\eqref{eq:alpharel} relates the ratio of the TRPMD and RPMD rates as a function of $\lambda$ with one parameter $\tilde \alpha_{\rm RP}$, and without requiring knowledge of the precise location of the optimal dividing surface $f^*(\bq)$. However, we can use general observations concerning which ring-polymer normal modes contribute to $f^*(\bq)$ to determine the likely sensitivity of the TRPMD rate to friction. Above the crossover temperature into deep tunnelling, defined by\cite{ric09}
\begin{align}
 \beta_c = \frac{2\pi}{\hbar \omega_b} \eql{xover}
\end{align}
where $\omega_b$ is the barrier frequency in the external potential $V(q)$, the optimal dividing surface is well approximated by the centroid\cite{ric09}
\begin{align}
 f^*(\bq) = \frac{1}{N}\smiNz q_i - \qdd. \eql{cen}
\end{align}
where $\qdd$ is the maximum in $V(q)$. As the centroid is not thermostatted (since $\bm{\Omega}_{00} = 0$), in this regime $\gamma_{\rm RP} = 0 = \tilde \alpha_{\rm RP}$ and we therefore predict from \eqr{alpharel} that the rate will be independent of $\lambda$, i.e. $k_{\rm TRPMD}(\beta) \simeq k_{\rm RPMD}(\beta)$.

Beneath the crossover temperature, the saddle point on the ring-polymer potential energy surface bends into the space of the first degenerate pair of normal modes.\cite{ric09,ric12} For symmetric systems, the optimal dividing surface is still the centroid expression in \eqr{cen} and (insofar as the reaction dynamics can be considered one-dimensional) $\tilde \alpha_{\rm RP} \simeq 0$, so $k_{\rm TRPMD}(\beta) \simeq k_{\rm RPMD}(\beta)$. 
 
For asymmetric reactions, the optimal dividing surface is now a function of both the centroid and first degenerate pair of normal modes (which are thermostatted)\cite{ric09}, and we expect $\tar > 0$. From \eqr{alpharel} the TRPMD rate will decrease linearly with $\lambda$ for small $\lambda$, for large friction as $\lambda^{-1}$, and the ratio of the TRPMD to RPMD rates to be a convex function of $\lambda$. This behaviour would also be expected for symmetric reactions beneath the second crossover temperature where the optimal dividing surface bends into the space of the second degenerate pair of normal modes.\cite{ric09} In all cases one would expect that increasing friction would either have no effect on the rate, or at sufficiently low temperatures cause it to decrease.

It should be stressed that \eqr{alpharel} is a considerable simplification of the TRPMD dynamics and is not expected to be reliable in systems where the ring polymer potential energy surface is highly anharmonic or skewed (such as F+H$_2$ investigated below). In fact, even for a one-dimensional system, the minimum energy path on the $N$-dimensional ring polymer potential energy surface shows a significant skew beneath the crossover temperature\cite{ric12}. The utility of \eqr{alpharel} lies in its simplicity and qualitative description of friction-induced recrossing.

\section{Results}
\label{sec:num}
We initially study the benchmark one-dimensional symmetric and asymmetric Eckart barriers before progressing to the multidimensional reactions H+H$_2$ (symmetric), D+MuH (quasisymmetrical), H+CH$_4$ (asymmetric, polyatomic) and F+H$_2$ (asymmetric, anharmonic).

\subsection{One-dimensional results}
The methodology for computation of TRPMD reaction rates is identical to that of RPMD\cite{lix13}, except for the thermostat attached to the internal normal modes of the ring polymer, achieved using the algorithm in Ref.~\onlinecite{cer10}. The Bennett-Chandler\cite{fre02} factorization was employed, and the same dynamics can be used for thermodymamic integration along the reaction co-ordinate (to calculate the QTST rate) as to propagate trajectories (to calculate the transmission coefficient).\cite{cer10,ros14} 

\begin{table*}[tb]
\caption{Dimensionless friction sensitivity parameter $\tar$ from \eqr{alpharel}, fitted by nonlinear least squares to simulation data.}
\label{tab:tar}
{\newcommand{\mc}[3]{\multicolumn{#1}{#2}{#3}}
\begin{tabular}{llccc|lccc} 
\hline
\hline
& \mc{3}{c}{ 1D Eckart barriers}  &  & \mc{4}{c}{Multidimensional reactions}\\
\cline{2-9}
\multirow{3}{*}{Symmetric} & $k_{\rm B}\beta/10^{-3}\textrm{K}^{-1}$  & 3 & 5 & 7 & $T$/K & 500 & 300 & 200 \\
\cline{2-9}
&  & $<$0.01 & 0.11 & 0.37 & H+H$_2$ &  0.01 & 0.16 & 0.45 \\
 & & &  &  & D+MuH & 0.20 & 0.45 & 0.71\\ \hline
\multirow{3}{*}{Asymmetric} &  $\beta$/a.u. & 4 & 8 & 12 &  $T$/K &  500  & 300 & 200 \\
\cline{2-9}
&  & 0.00 & 0.06 & 0.17 & H+CH$_4$  &  0.00 & 0.10 &  0.16 (250K) \\ 
&  &  &  &  & F+H$_2$ &  --0.01 & --0.01 & 0.00 \\ \hline\hline
\end{tabular}
}%
\end{table*}

We firstly examine the symmetric Eckart barrier\cite{eck30,man05ref},
\begin{align}
V(q)=V_0\ \text{sech}^2(q/a),  
\end{align}
and to facilitate comparison with the literature\cite{man05ref,ric09,zha14com}, use parameters to model the H+H$_2$ reaction: $V_0 = 0.425$eV, $a = 0.734a_0$, and $m=1061m_e$, leading to a crossover temperature of $k_{\rm B}\beta_c = 2.69\times 10^{-3}\textrm{K}^{-1}$. The centroid reaction co-ordinate of \eqr{cen} was used throughout. Results for a variety of temperatures and values of friction parameter $\lambda$ are presented in \figr{sym}, and values of $\tar$ obtained by nonlinear least squares in Table~\ref{tab:tar}. 

\begin{figure}[tb]
 \includegraphics[width=0.5\textwidth]{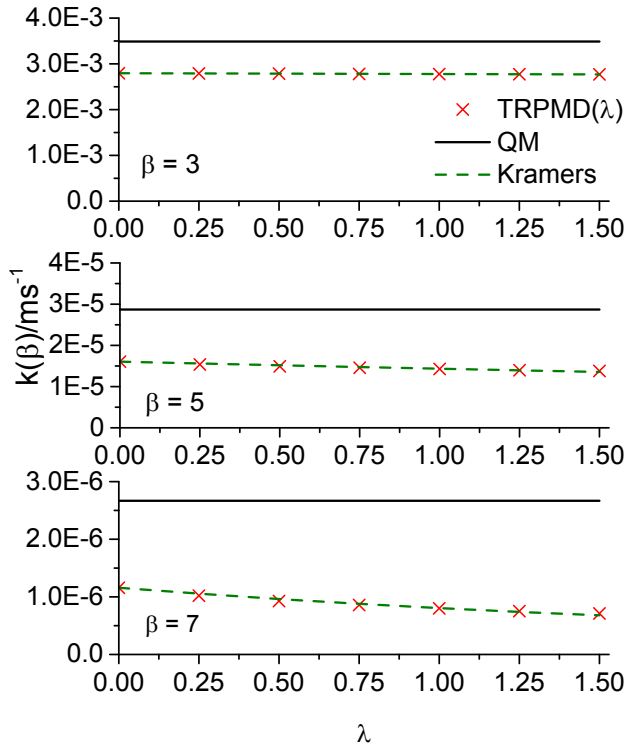}
 \caption{Results for the symmetric Eckart barrier, showing the TRPMD result as a function of $\lambda$ (red crosses), fitted Kramers curve (green dashes) and quantum result (black line). $\beta$ is quoted in units of $k_{\rm B}^{-1}10^{-3}\textrm{K}^{-1}$ and the crossover temperature is $k_{\rm B}\beta_c = 2.69\times 10^{-3}\textrm{K}^{-1}$.}
 \figl{sym}
\end{figure}

Slightly beneath the crossover temperature ($k_{\rm B}\beta_c = 3\times 10^{-3}\textrm{K}^{-1}$), the TRPMD rate is indepedent of the value of friction ($\tar = 0$), as predicted by Kramers theory. Some sensitivity to $\lambda$ is seen before twice the crossover temperature, which is likely to be a breakdown of the one-dimensional assumption of Kramers theory; while the centroid is the optimal dividing surface, the minimum energy path bends into the space of the (thermostatted) lowest pair of normal modes\cite{ric12}. Beneath twice the crossover temperature the friction parameter has a significant effect on the rate, as to be expected from the second degenerate pair of normal modes becoming part of the optimal dividing surface\cite{ric09}. The functional form of $\chi_{\lambda}(\beta)$ is also in accordance with the predictions of Kramers theory, monotonically decreasing as $\lambda$ rises, and being a convex function of $\lambda$. 

Since RPMD underestimates the rate for this symmetric reaction (and many others\cite{ric09}), adding friction to RPMD decreases its accuracy in approximating the quantum rate for this system.

The asymmetric Eckart barrier is given by\cite{man05ref}
\begin{align}
 V(q) = \frac{A}{1+e^{-2q/a}} + \frac{B}{\cosh^2(q/a)}
\end{align}
where $A = -18/\pi$, $B = 13.5/\pi$ and $a = 8/\sqrt{3\pi}$ in atomic units ($\hbar = k_{\rm B} = m = 1$), giving a crossover temperature of $\beta_c = 2\pi$. To facilitate comparison with previous literature\cite{man05ref,ric09,pol98,jan00} the results are presented in \figr{asymlam} as the ratio
\begin{align}
 c(\beta) = \frac{k(\beta)}{k_{\rm clas}(\beta)} \eql{cbeta}
\end{align}
and $\tar$ values in Table~\ref{tab:tar}.

\begin{figure}[tb]
 \includegraphics[width=0.5\textwidth]{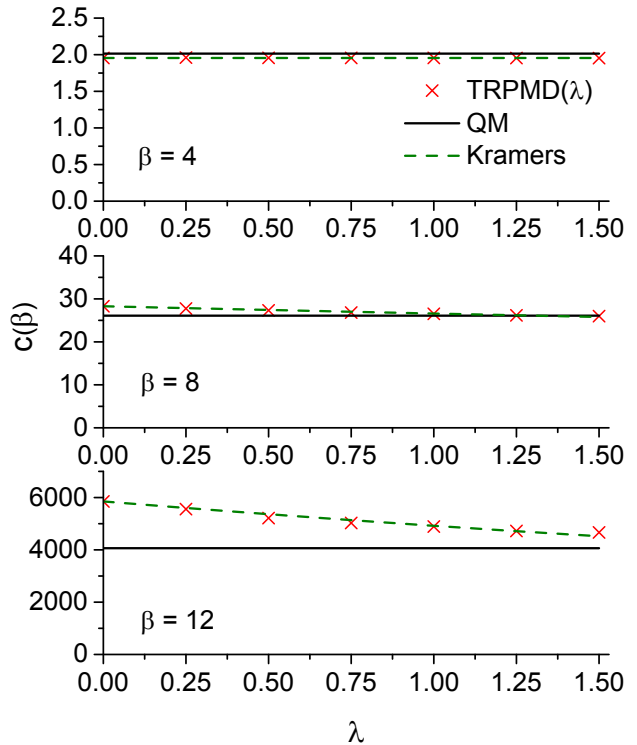}
 \caption{Results for the asymmetric Eckart barrier quoted as $c(\beta)$ [\eqr{cbeta}], showing the TRPMD result as a function of $\lambda$ (red crosses), fitted Kramers curve (green dashes) and quantum result (black line). The crossover temperature is $\beta_c = 2\pi$a.u.}
 \figl{asymlam}
\end{figure}

Above the crossover temperature, TRPMD is invariant to the value of the friction parameter, and beneath the crossover temperature, increasing $\lambda$ results in a decrease in the rate, such that TRPMD is closer to the exact quantum result than RPMD for all $\lambda>0$ in this system. The decrease in the TRPMD rate with $\lambda$ is qualitatively described by the crude Kramers approximation (see \figr{asymlam}), and it therefore seems that the improved accuracy of TRPMD could be a fortuitous cancellation between the overestimation of the quantum rate by QTST, and the friction-induced recrossing of the optimal dividing surface by TRPMD trajectories. There is no particular \emph{a priori} reason to suppose that one value of $\lambda$ should provide superior results; from \figr{asymlam}, at $\beta = 8$ a friction parameter of $\lambda = 1.25$ causes TRPMD to equal the quantum result to within graphical accuracy, whereas at $\beta=12$ this value of friction parameter causes overestimation of the rate, and further calculations (not shown) show that $\lambda=5$ is needed for TRPMD and the quantum rates to agree.


The numerical results also show a slightly higher curvature in $k_{\rm TRPMD}(\beta)$ as a function of $\lambda$ than \eqr{alpharel} would predict, suggesting that the TRPMD rate reaches an asymptote at a finite value, rather than at zero as the Kramers model would suggest. We suspect this is a breakdown of one-dimensional Kramers theory, since in the $\lambda \to \infty$ limit the system can still react via the unthermostatted centroid co-ordinate, but may have to surmount a higher barrier on the ring polymer potential energy surface.

\begin{figure}[tb]
 \includegraphics[width=0.5\textwidth]{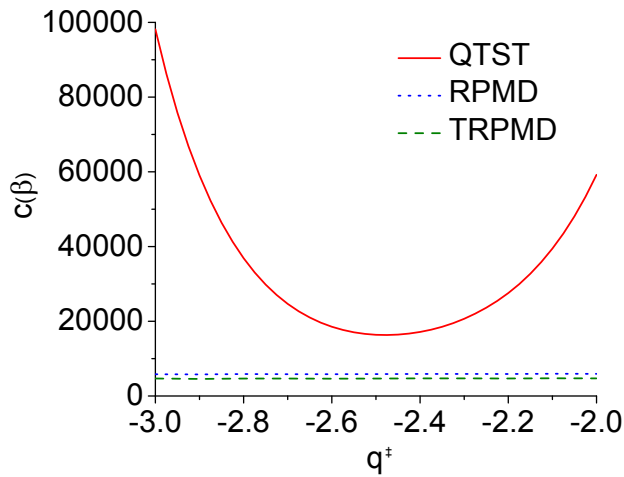}
 \caption{TRPMD (green dashes), RPMD (blue dots) and QTST (centroid dividing surface, red line) rates for the asymmetric Eckart barrier at $\beta = 12$, as a function of the dividing surface $\qdd$.}
 \figl{dsinv}
\end{figure}

We then investigate the effect of changing the location of the centroid dividing surface on the TRPMD rate. RPMD is already known to be invariant to the location of the dividing surface\footnote{This proof\cite{man05ref} was originally for a gas-phase system and a centroid dividing surface, but this can easily be extended to the condensed phase provided there is the necessary separation of timescales\cite{cha78}.}, and we therefore choose a system for which TRPMD and RPMD are likely to differ the most, namely a low-temperature, asymmetric system where there is expected to be significant involvement of the thermostatted lowest degenerate pair of normal modes in crossing the barrier. The asymmetric Eckart barrier at $\beta=12$ is therefore used as a particularly harsh test, with the result plotted in \figr{dsinv}. Although the centroid-density QTST result varies by almost a factor of six across the range of dividing surfaces considered ($-3\le \qdd \le -2$a.u.), both the TRPMD and RPMD rates are invariant to the location of the dividing surface. We also observe that, even with the optimal dividing surface, centroid-density QTST significantly overestimates the exact rate\cite{man05ref,jan00}.

\subsection{Multidimensional results}
\begin{table*}
\caption{\scriptsize Input parameters for the TRPMD calculations on the H + H$_2$, D + MuH, and F + H$_2$ reactions. 
The explanation of the format of the input file can be found in the RPMDrate code manual 
(see Ref. 82 and \href{http://www.mit.edu/~ysuleyma/rpmdrate}{http://www.mit.edu/~ysuleyma/rpmdrate}).}
\label{tab1}
\begin{center}
\begin{tabular}{llllll} 
\hline
\hline
Parameter & \multicolumn{4}{c}{Reaction} & Explanation \\
\cline{2-6}
 & H + H$_2$ & D + MuH & F + H$_2$  & H + CH$_4$ \\
\hline
\multicolumn{6}{l}{Command line parameters} \\ 
\hline
\ttfamily{Temp}         & \multicolumn{3}{l}{200; 300; 500}     & 250; 300; 500 &      Temperature (K)  \\
\ttfamily{Nbeads}       & 128 & 512 & 384 (200 K) &  192 (250 K) &Number of beads in the TRPMD calculations  \\
  & & & 256 (300 K) & 128 (300 K) \\
  & & & 64 (500 K) &  64 (500 K)\\	
\hline
\multicolumn{6}{l}{Dividing surface parameters} \\ 
\hline
$R_\infty $  & 30   & 30  & 30  & 30 & Dividing surface $s_1$ parameter ($a_0$) \\
$N_{\rm bonds}$       & 1    & 1   & 1   &  1 & Number of forming and breaking bonds \\
$N_{\rm channel}$     & 2    & 1   & 2   &  4 & Number of equivalent product channels \\
\hline
\multicolumn{5}{l}{Thermostat options} \\ 
\hline
\ttfamily{thermostat}   & \multicolumn{4}{l}{'GLE/Andersen'}  &  Thermostat for the QTST calculations \\
\ttfamily{$\lambda $}   & \multicolumn{4}{l}{0; 0.25; 0.5; 0.75; 1.0; 1.5} &  Friction coefficient for the recrossing factor calculations \\
\hline
\multicolumn{5}{l}{Biased sampling parameters} \\ 
\hline
$N_{\rm windows}$  & 111    & 111     & 111     & 111         & Number of windows \\
$\xi _1$            & -0.05  & -0.05   & -0.05  & -0.05   & Center of the first window \\
$d\xi $              & 0.01   & 0.01   & 0.01   & 0.01     & Window spacing step      \\ 
$\xi _N$             & 1.05   & 1.05   & 1.05   & 1.05  & Center of the last window \\
$dt$  & 0.0001 & 0.0001 & 0.0001 & 0.0001 &  Time step (ps) \\
$k_i$  &  2.72  & 2.72  & 2.72  & 2.72  & Umbrella force constant ((T/K) eV) \\
$N_{\rm trajectory}$ & 200 & 200 & 200 &  200 & Number of trajectories \\ 
$t_{\rm equilibration}$ & 20  & 20  & 20  &  20  & Equilibration period (ps)  \\
$t_{\rm sampling}$ &  100 & 100   & 100  &  100  & Sampling period in each trajectory (ps) \\
$N_i$   & $2\times 10^8$ & $2\times 10^8$ & $2\times 10^8$  & $2\times 10^8$  &  Total number of sampling points \\
\hline
\multicolumn{6}{l}{Potential of mean force calculation} \\ 
\hline
$\xi _0$            & -0.02  & -0.02  & -0.02  & -0.02  & Start of umbrella integration  \\
$\xi ^{\ddagger}$             & 1.0000$*$  &0.9912 (200 K)$*$   & 0.9671 (200 K)$*$  &  1.0093 (250 K)$*$  & End of umbrella integration \\
 & & 0.9904 (300 K)$*$ &  0.9885 (300 K)$*$ &  1.0074 (300 K)$*$  & \\
& & 0.9837 (500 K)$*$  &  0.9947 (500 K)$*$ & 1.0026 (500 K)$*$ & \\

$N_{\rm bins}$             & 5000  & 5000  & 5000  & 5000  &  Number of bins \\
\hline
\multicolumn{6}{l}{Recrossing factor calculation} \\ 
\hline
$dt$  & 0.0001 & 0.00003 & 0.0001 &  0.0001 &Time step (ps) \\
$t_{\rm equilibration}$  & 20  & 20  & 20  &  20  & Equilibration period (ps) in the constrained (parent)\\ 
 & & & &  & trajectory \\
$N_{\rm totalchild}$  & 100000  & 100000  & 500000   & 500000   & Total number of unconstrained (child) trajectories  \\
$t_{\rm childsampling}$  & 20  & 20  & 20  &  20  & Sampling increment along the parent trajectory  (ps)  \\
$N_{\rm child}$  & 100  & 100  & 100  &  100  & Number of child trajectories per one   \\
  & & & & & initially constrained configuration \\
$t_{\rm child}$  & 0.05  & 0.2  & 0.2  &   0.1  &Length of child trajectories (ps)  \\
\hline
\hline
\end{tabular}\\
\end{center}
$*$ Detected automatically by RPMDrate. \\
\end{table*}

The results are calculated using adapted RPMDrate code\cite{sul13}, with details summarized in Table~\ref{tab1}. In the calculations reported below we used the potential energy surface developed by Boothroyd et al.\ (BKMP2 PES) for H+H$_2$ and D+MuH,\cite{boo96} the Stark–Werner (SW) potential energy surface for F+H$_2$,\cite{sta96} and the PES-2008 potential energy surface developed by Corchado et al.\ for H+CH$_4$.\cite{cor09} The computation of the free energy was achieved using umbrella integration\cite{kas05,kas06} with TRPMD and checked against standard umbrella integration with an Andersen thermostat\cite{and83}.

\begin{figure}[tb]
 \includegraphics[width=0.5\textwidth]{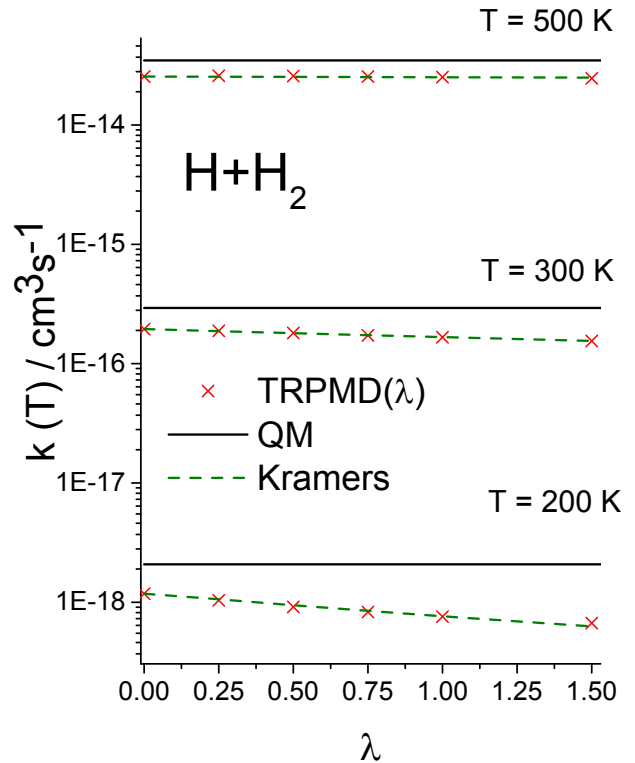}
 \caption{Results for the H+H$_2$ reaction as a function of $\lambda$. Kramers is the fitted Kramers curve (see text). The crossover temperature is 345K.}
 \figl{hhh}
\end{figure}

H+H$_2$ represents the simplest atom-diatom scattering reaction and has been the subject of numerous studies\cite{sul13hh2,col09,jan14simp}. The PES is symmetric and with a relatively large skew angle (60\textdegree), and a crossover temperature of 345K. The results in \figr{hhh} show that the rate is essentially invariant to the value of $\lambda$ above the crossover temperature. At 300K there is a slight decrease in the rate with increasing friction from 0 to 1.5 ($\sim $25 $\%$), and this is far more pronounced at 200K where the $\lambda = 1.5$ result is almost half that of the $\lambda = 0$ (RPMD) result. 

\begin{figure}[tb]
 \includegraphics[width=0.5\textwidth]{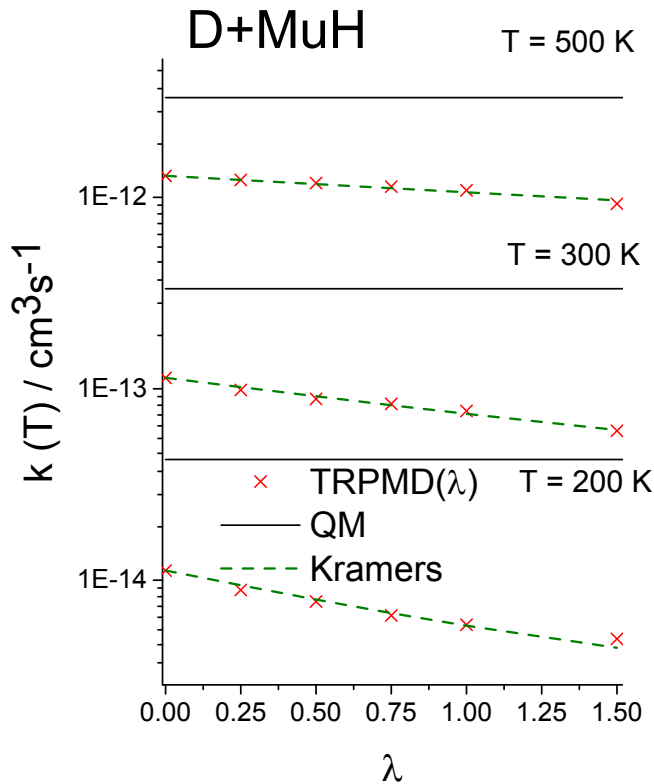}
 \caption{As for \figr{hhh}, but for the D+MuH reaction with a high crossover temperature of 860K.}
 \figl{dmuh}
\end{figure}

D+MuH is ``quasisymmetrical'' since DMu and MuH have very similar zero-point energies, and one would therefore expect the RPMD rate to underestimate the exact quantum rate\cite{per14stress}. Since it is Mu-transfer the crossover temperature is very high (860 K) and therefore this reaction can be considered as a stress test for the deep tunneling regime. The results in \figr{dmuh} show that friction in the TRPMD dynamics causes further underestimation of the rate,  especially at low temperatures; for $\lambda = 1.5$ at 200K, TRPMD underestimates the exact quantum rate by almost an order of magnitude, and
even at 500K it decreases by $\sim$40\% over the range of $\lambda $ explored here.  

\begin{figure}[tb]
 \includegraphics[width=0.5\textwidth]{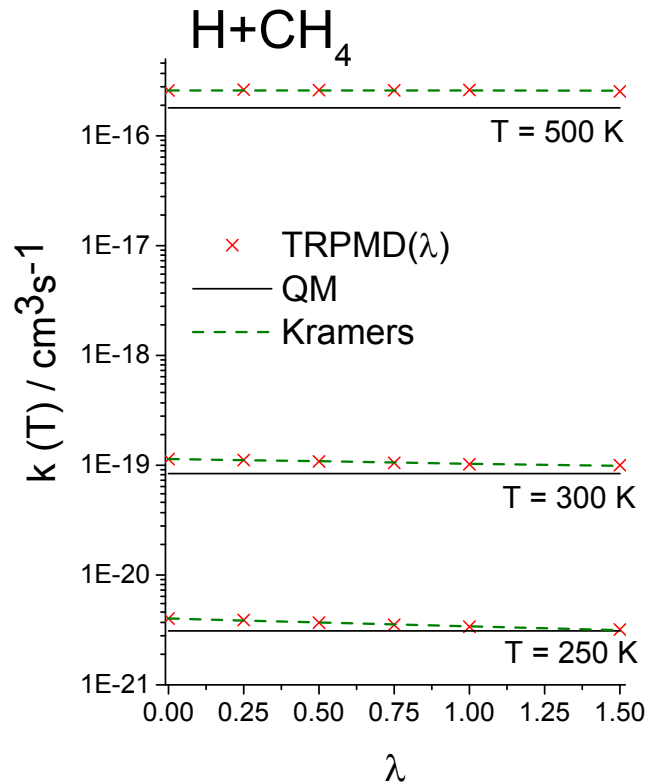}
 \caption{Results for the H+CH$_4$ reaction. The crossover temperature is 341K.}
 \figl{hch4}
\end{figure}

As an example of a typical asymmetric reaction, results for H+CH$_4$ are plotted in \figr{hch4}, which has a crossover temperature of 341K. RPMD is well-known to overestimate the quantum rate for this system at low temperatures.\cite{sul11} \figr{hch4} shows that above the crossover temperature (500K) the  friction parameter has a negligible effect on the rate. As the temperature is decreased below the crossover temperature (300K and 250K), the friction induces more recrossings of the dividing surface and, as a result, the TRPMD rate approaches the exact quantum rate with increasing the friction parameter. 

\begin{figure}[tb]
 \includegraphics[width=0.5\textwidth]{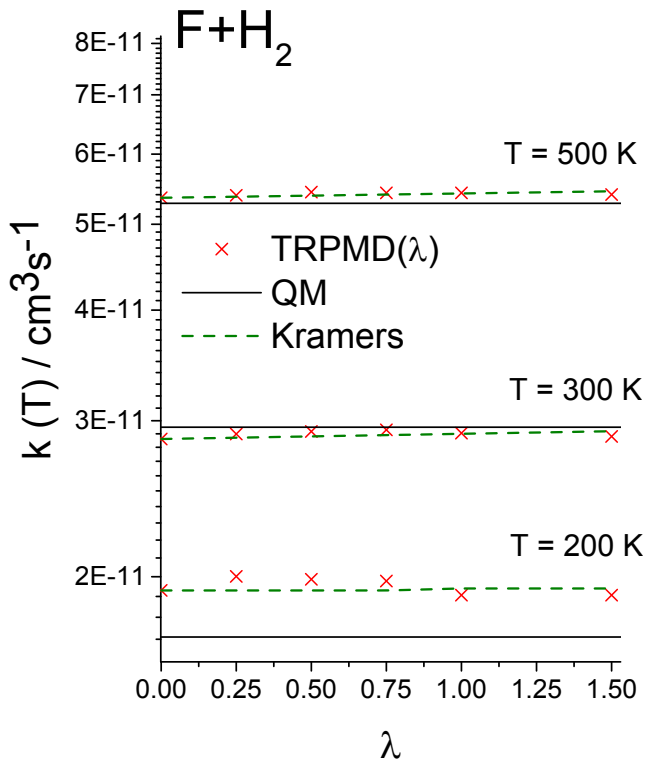}
 \caption{Results for the anharmonic and asymmetric F+H$_2$ reaction with a crossover temperature of 264K.}
 \figl{fhh}
\end{figure}

Thus far, Kramers theory has been surprisingly successful at qualitatively explaining the behaviour of the TRPMD rate with increasing friction. Present results would suggest that TRPMD would therefore improve upon RPMD for all asymmetric reactions, where RPMD generally overestimates the rate beneath crossover\cite{ric09,sul13hh2}. We then examine another prototypical asymmetric reaction, F+H$_2$, with a low crossover temperature of 264K. \figr{fhh} shows that at 500K and 300K, the TRPMD rate is in good agreement with the quantum result, but increases very slightly with $\lambda$ causing a spurious small negative value of $\tar$ in Table~\ref{tab:tar}. Beneath crossover, at 200K the rate is virtually independent of lambda, apart from a very slight increase around $\lambda = 0.5$. Consequently, TRPMD fares no better than RPMD for this system, contrary to the H+CH$_4$ results and the predictions of Kramers theory. This is likely attributable to a highly anharmonic and exothermic energy profile, and a very flat saddle point in ring-polymer space\cite{sta96,ste10}.

As can be seen from the graphs, the simple Kramers prediction is in surprisingly good qualitative agreement with the numerical results (apart from F+H$_2$ beneath crossover), even for the multidimensional cases, which is probably attributable to those reactions being dominated by a significant thermal barrier which appears parabolic on the ring-polymer potential energy surface, meaning that the one-dimensional Kramers model is adequate for capturing the friction-induced recrossing. In Table~\ref{tab:tar} the $\tar$ values, fitted to the numerical data, show that for a given reaction $\tar\simeq 0$ above the crossover temperature, and beneath the crossover temperature $\tar$ increases as the temperature is decreased. This can be qualitatively explained as the optimal dividing surface becoming more dependent on the thermostatted higher normal modes as the temperature is lowered\cite{ric09}. Not surprisingly, the highest value of $\tar$ is observed for the highly quantum mechanical D+MuH reaction at 200K with $\tar = 0.71$. This is beneath one quarter of the crossover temperature, and one would therefore expect that friction would have a very significant effect on the rate. 

\section{Conclusions}
\label{sec:con}
In this paper we have, for the first time, applied Thermostatted Ring Polymer Molecular Dynamics (TRPMD) to reaction rate theory. Regardless of the applied friction, the long-time limit of the TRPMD flux-side time-correlation function (and therefore the TRPMD rate) is independent of the location of the dividing surface, and its short-time limit is equal to rigorous QTST\cite{hel13,alt13,hel13unique,ros14}. In section~\ref{ssec:kram} we use Kramers theory \cite{kra40} to predict that, above the crossover temperature, the RPMD and TRPMD rates will be similar, and beneath crossover the TRPMD rate for asymmetric systems will decrease with $\lambda$, and the same effect should be observed for symmetric systems beneath half the crossover temperature.

TRPMD rate theory has then been applied to the standard one-dimensional model systems of the symmetric and asymmetric Eckart barriers, followed by the bimolecular reactions H+H$_2$, D+MuH, H+CH$_4$ and F+H$_2$. For all reactions considered, above the crossover temperature the TRPMD rate is virtually invariant to the value of $\lambda$ and therefore almost equal to RPMD, as predicted by Kramers theory. Beneath the crossover temperature, most asymmetric reactions show a decrease in the TRPMD rate as $\lambda$ is increased, and in qualitative agreement with the Kramers prediction in \eqr{alpharel}. A similar trend is observed for symmetric reactions, which also show some diminution in the rate with increasing friction above half the crossover temperature ($\beta_c < \beta < 2 \beta_c$), probably due to the skewed ring-polymer PES causing a breakdown in the one-dimensional assumption of Kramers theory. For the asymmetric and anharmonic case of F+H$_2$, beneath the crossover temperature there is no significant decrease in the rate with increased friction, illustrating the limitations of Kramers theory.

These results mean that beneath the crossover temperature TRPMD will be a worse approximation to the quantum result than RPMD for symmetric and quasisymmetrical systems (where RPMD underestimates the rate\cite{sul13hh2,ric09}), and TRPMD will be closer to the quantum rate for asymmetric potentials (where RPMD overestimates the rate). However, the apparent increase in accuracy for asymmetric systems appears to be a cancellation of errors from the overestimation of the quantum rate by RPMD which is then decreased by the friction in the non-centroid normal modes of TRPMD, and there is no a priori reason to suppose that one effect should equal the other for any given value of $\lambda$. 

Although the above results do not advocate the use of TRPMD rate theory as generally being more accurate than RPMD, TRPMD rate calculation above the crossover temperature may be computationally advantageous in complex systems due to more efficient sampling of the ring-polymer phase space by TRPMD trajectories than RPMD trajectories\footnote{M.~Rossi and D.~E.~Manolopoulos, private communication, (2015).}. TRPMD may therefore provide the same accuracy as RPMD rate calculation at a lower computational cost, and testing this in high-dimensional systems where RPMD has been successful, such as complex-forming reactions\cite{sul14,lix14,hic15,per14}, surface dynamics\cite{sul12}, and enzyme catalysis\cite{boe11} would be a useful avenue of future research.

Future work could also include non-adiabatic systems\cite{hel11,ana13,ana10,men14,ric13,ric14,men11,kre13}, applying a thermostat to the centroid to model a bath system\cite{man05che}, and generalizations to non-Markovian friction using Grote-Hynes theory\cite{gro80}. 

In closing, present results suggest that TRPMD can be used above the crossover temperature for thermally activated reactions, and beneath crossover further testing is required to assess its utility for asymmetric systems.

\section{Acknowledgements}
TJHH acknowledges a Research Fellowship from Jesus College, Cambridge, and helpful comments on the manuscript from Stuart Althorpe. YVS acknowledges support via the Newton International Alumni Scheme from the Royal Society. YVS also thanks the European Regional Development Fund and the Republic of Cyprus for support through the Research Promotion Foundation (Project Cy-Tera NEA $\Gamma\Pi$O$\Delta$OMH/$\Sigma$TPATH/0308/31).


\appendix
\section{Detailed Balance}
\label{ap:db}
For a homogeneous Markov process such as TRPMD for which negative time is not defined\cite{zwa01}, detailed balance is defined as\cite{gar09}
\begin{align}
 \mathcal{P}&(\bp', \bq',t|\bp,\bq,0) \rho_s(\bp,\bq) \no\\
 &= \mathcal{P}(-\bp, \bq,t|-\bp',\bq',0) \rho_s(\bp',\bq') \eql{db}
\end{align}
where $\rho_s(\bp,\bq)=e^{-\betaN H_{N}(\bp,\bq)}$ is the stationary distribution and $\mathcal{P}(\bp', \bq' ,t|\bp,\bq,0)$ is the conditional probability that a ring polymer will be found at point $(\bp',\bq')$ at time $t$, given that is was at $(\bp,\bq)$ at time $t=0$. 

To demonstrate that \eqr{db} is statisfied, we rewrite the Fokker-Planck operator \eqr{fpo} as
\begin{align}
 \mathcal{A}_N = & - \smjNz \left(\ddp{}{q_j} a(\bp,\bq)_j + \ddp{}{p_j} b(\bp,\bq)_j\right) \no\\
 & + \frac{1}{2} \smjNz \sum_{j'=0}^{N-1} \ddp{}{p_j}\ddp{}{p_{j'}}  C(\bp,\bq)_{jj'} \eql{sumsum}
\end{align}
where the vectors ${\bf a}(\bp,\bq) = \bp/m$, ${\bf b}(\bp,\bq) = -U_{N}(\bq) \ola \nabla_{\bq} - \bm{\Gamma} \cdot \bp$ and the matrix ${\bf C}(\bp,\bq) = 2m\bm{\Gamma}/\betaN$. Note that the derivatives in \eqr{sumsum} act on ${\bf a}(\bp,\bq)$, ${\bf b}(\bp,\bq)$ or ${\bf C}(\bp,\bq)$ and whatever follows them which is acted upon by $\mathcal{A}_N$.

The necessary and sufficient conditions for detailed balance [\eqr{db}] to hold, in addition to $\rho_s(\bp,\bq)$ being a stationary distribution, are then given by\cite{gar09} 
\begin{align}
 {\bf a}(-\bp,\bq)\rho_s(\bp,\bq)  = & - {\bf a}(\bp,\bq) \rho_s(\bp,\bq)  \eql{con1}\\
 -{\bf b}(-\bp,\bq)^T \rho_s(\bp,\bq) = & -{\bf b}(\bp,\bq)^T \rho_s(\bp,\bq) \no\\
& + \nabla_{\bf p} \cdot {\bf C}(\bp,\bq)   \rho_s(\bp,\bq) \eql{con2}\\
 {\bf C}(-\bp,\bq) = & {\bf C}(\bp,\bq) \eql{con3}
\end{align}
Condition \eqr{con1} is trivially satisfied. Provided that the friction matrix is even w.r.t.~momenta (satisfied here as $\bm{\Gamma}$ is not a function of $\bp$) \eqr{con3} will be satisfied. \eqr{con2} becomes
\begin{align}
 ({\bm \Gamma} \cdot \bp)^T \rho_s(\bp,\bq) = -\frac{m}{\betaN} \nabla_{\bp} \cdot \bm{\Gamma} \rho_s(\bp,\bq)
\end{align}
which is satisfied with $\rho_s(\bp,\bq) = e^{-\betaN H_{N}(\bp,\bq)}$ and the friction matrix used here.

Given that \eqr{db} is satisfied, for an arbitrary correlation function one can then show
\begin{widetext}
\begin{align}
 C_{AB}^{\rm TRPMD}(t) = & \tphN \int d\bp \int d\bq \int d\bp' \int d\bq'\ e^{-\betaN H_N(\bp,\bq)} A(\bp,\bq) \mathcal{P}(\bp', \bq',t|\bp,\bq,0) B(\bp',\bq') \\
 = & \tphN \int d\bp \int d\bq \int d\bp' \int d\bq'\ e^{-\betaN H_N(\bp,\bq)} A(-\bp',\bq') \mathcal{P}(\bp', \bq',t|\bp,\bq,0) B(-\bp,\bq)
\end{align}
\end{widetext}
and for the Langevin trajectories considered here, which are continuous but not differentiable, this means
\begin{align}
 C_{AB}^{\rm TRPMD}(t) = & \tphN \int d\bp \int d\bq \ e^{-\betaN H_N(\bp,\bq)}\no\\
 & \qquad \times  A(\bp,\bq) \bar B(\bp_t,\bq_t) \\
 = & \tphN \int d\bp \int d\bq  \ e^{-\betaN H_N(\bp,\bq)}\no\\
 & \qquad \times \bar A(-\bp_t,\bq_t) B(-\bp,\bq)
\end{align}
where $\bq_t\equiv \bq_t(\bp,\bq,t)$ is the vector of positions stochastically time-evolved according to \eqsr{dynp}{dynq}, and
\begin{align}
  \bar B(\bp_t,\bq_t) =  \int d\bp' \int d\bq'\ \mathcal{P}(\bp', \bq',t|\bp,\bq,0)B(\bp',\bq')
\end{align}
with $\bar A(\bp_t,\bq_t)$ similarly defined.

\section{Independence of $k_{\rm TRPMD}(\beta)$ to the dividing surface location}
\label{ap:ind}
We use a similar methodology to that which Craig and Manolopoulos employed for RPMD\cite{man05ref}, and give the main steps here. 
We firstly differentiate the side-side correlation function in \eqr{css} w.r.t.\ the location of the dividing surface $\qdd$ (or any other parameter specifying the nature of the dividing surface), giving
\begin{align}
 \dd{}{\qdd}C_{\rm ss}(t) = & \tphN \int d\bp \int d\bq \ e^{-\betaN H_N(\bp,\bq)} \no\\
 & \times \ddp{\fq}{\qdd}\left\{\dfq h[f(\bq_t)] + h[\fq] \delta[f(\bq_t)] \right\}. \eql{dcssdq}
\end{align}
Since TRPMD dynamics obeys detailed balance (as shown in appendix~\ref{ap:db}), and the dividing surface is only a function of position, the second term on the RHS of \eqr{dcssdq} is identical to the first,
\begin{align}
 \dd{}{\qdd}C_{\rm ss}(t) = & \frac{2}{(2\pi\hbar)^{N}} \int d\bp \int d\bq \ e^{-\betaN H_N(\bp,\bq)} \no\\
 & \qquad \times \ddp{\fq}{\qdd}\dfq h[f(\bq_t)]. \eql{dcssdq2}
\end{align}
Differentiation of \eqr{dcssdq2} w.r.t.\ time using \eqr{fpa}, and relating the side-side and flux-side functions using \eqr{cssfs}, yields
\begin{align}
 \dd{}{\qdd}C_{\rm fs}(t) = & -\frac{2}{(2\pi\hbar)^{N}} \int d\bp \int d\bq \ e^{-\betaN H_N(\bp,\bq)}\no\\
 & \qquad \times \ddp{\fq}{\qdd}\dfq \delta[f(\bq_t)] S_N(\bp_t,\bq_t). \eql{dcfsdqdd}
\end{align}
Equation~\eqref{eq:dcfsdqdd} corresponds to a trajectory commencing at the dividing surface at time zero and returning to it at time $t$ with non-zero velocity $S_N(\bp_t,\bq_t)$. At finite times while there is recrossing of the barrier, there will be trajectories satisfying these conditions, but after the plateau time when no trajectories recross the barrier [cf.\ \eqr{cff}], these conditions are clearly not satisfied, and the rate will be independent of the location of the dividing surface.\cite{man05ref}

This proof is valid for any friction matrix which satisfies the detailed balance conditions of appendix~\ref{ap:db}, and does not require the presence of ring-polymer springs in the potential, so is valid for \emph{any} classical-like reaction rate calculation using Langevin dynamics.

\end{document}